\documentclass[twocolumn,longbibliography,prl,tightenlines,superscriptaddress]{revtex4-2}

\usepackage{hyperref}
\usepackage{amsmath,amssymb}
\usepackage[utf8]{inputenc} 			
\usepackage[T1]{fontenc}				
\usepackage{graphicx}
\usepackage{float}
\usepackage[usenames,dvipsnames]{xcolor}
\usepackage{tikz}

\renewcommand{\vec}[1]{\mathbf{#1}}

\graphicspath{{./Image/}}

\usepackage{lipsum}

\usepackage{tikz}

\begin{document}

\title{Understanding the Excitability Phase Transition of Polar Flocks}

\title{How Continuous Symmetry Stabilizes the Ordered Phase of Polar Flocks}

\author{Omer Granek}
\affiliation{Leinweber Institute for Theoretical Physics \& Kadanoff Center for Theoretical Physics, University of Chicago, 933 E 56th St, Chicago, Illinois 60637, USA}

\author{Hugues Chat\'{e}}
\affiliation{Service de Physique de l'Etat Condens\'e, CEA, CNRS Universit\'e Paris-Saclay, CEA-Saclay, 91191 Gif-sur-Yvette, France}
\affiliation{Computational Science Research Center, Beijing 100094, China}
\affiliation{Sorbonne Universit\'e, CNRS, Laboratoire de Physique Th\'eorique de la Mati\`ere Condens\'ee, 75005 Paris, France}

\author{Yariv Kafri}
\affiliation{Department of Physics, Technion – Israel Institute of Technology, Haifa 32000, Israel}

\author{Sunghan Ro}
\affiliation{Department of Physics, Harvard University, Cambridge, Massachusetts 02138, USA}

\author{Alexandre Solon}
\affiliation{Sorbonne Universit\'e, CNRS, Laboratoire de Physique Th\'eorique de la Mati\`ere Condens\'ee, 75005 Paris, France}

\author{Julien Tailleur}
\affiliation{Department of Physics, Massachusetts Institute of Technology, Cambridge, Massachusetts 02139, USA}

\begin{abstract}
  We study the stability of the ordered phase of compressible polar flocks against
  the nucleation of counter-propagating droplets, using a combination of analytical theory, microscopic and hydrodynamic simulations. For discrete-symmetry flocks,
  such droplets are known to always grow and propagate, making the ordered
  phase metastable. 
  We explain how, on the contrary, continuous symmetry can stabilize the
  ordered phase at small enough noise by destabilizing the leading edge of
  growing droplets. 
  Flocking models with continuous symmetries thus have a lower critical dimension than their discrete-symmetry counterparts, in contrast to equilibrium physics.
\end{abstract}

\maketitle

In equilibrium, 
systems that break a discrete symmetry have gapped lowest
energy excitations, while those that break a continuous symmetry
  have gapless ones~\cite{Kardar2007StatisticalFields}. An important
  consequence 
  is that a discrete symmetry can be broken
  down to a lower critical dimension $d_c=1+\epsilon$, while $d_c=2$
  for a continuous symmetry, as captured by the
  Hohenberg-Mermin-Wagner
  theorem~\cite{mermin_absence_1966,hohenberg_existence_1967}.  
  The
  underlying intuition is that Goldstone modes, since they stem from small field
  variations, are more easily excited than the nonlinear domain
  walls required to destroy order in discrete-symmetry systems.

\begin{figure}[t]
    \centering
    \includegraphics[width=\linewidth]{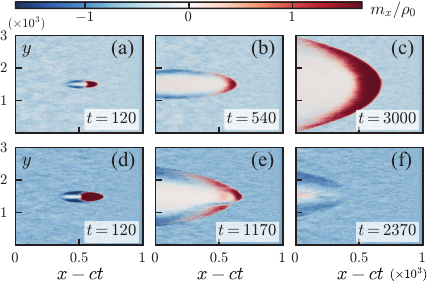}
    \caption{Fate of localized counter-propagating perturbations introduced in the ordered phase of the active XY model. 
    {\bf (a-c):} At $T=0.34$,  a ballistically growing droplet is observed. 
     {\bf (d-f):} At $T=0.31$, the perturbation eventually
      evaporates.
      (Initial circular perturbation of radius $20$ for $T = 0.34$ and $60$ for $T = 0.31$ with
      $\rho_d=10^2 \rho_0$, 
      $\rho_0=3$, $D=1$, $v\simeq0.55$, $\gamma=0.1$, so that ${\rm Pe}=3.0$. System size: $L_x = 1000, ~L_y = 3000$.) }
    \label{fig:droplets}
\end{figure}

  Compressible polar flocks offer a clear departure from this equilibrium
  picture: spin waves are unable to destroy orientational order in collections of 
  self-propelled particles aligning locally their velocity, 
even for two-dimensional systems with continuous rotational symmetry~\cite{toner1995long,Toner1998FlocksFlocking,toner2005hydrodynamics,Toner2012ReanalysisFlocks,Chate2024DynamicFlocks,jentsch_new_2024}.
  Recent results actually show an even stronger contrast with equilibrium.
  Discrete-symmetry flocks, in
  which particle self-propulsion is confined along a single
  dimension, exhibit no long-range order in any dimension~\cite{Benvegnen2023}. This is due to nonlinear
  excitations that cause
  droplets of the minority phase to grow ballistically and proliferate
  in the thermodynamic limit. On the contrary, numerical results on continuous-symmetry compressible flocks
  have shown that such growing droplets can exist but they disappear at low enough noise~\cite{Codina2022SmallFlock}. 
  The ordered phase of vectorial flocks thus seems more robust than that of scalar ones. This challenges our equilibrium intuition and currently lacks a theoretical explanation.
  
In this Letter, we uncover the mechanism that makes the existence of a continuous symmetry stabilize the ordered phase of compressible polar flocks. Using the active XY model 
and the associated mean-field continuum description, we study the stability of 
counter-propagating
droplets. We show that the presence of a continuous symmetry favors the emergence of a transverse polarization at their leading front
that tends to tear them apart. This effect is opposed by the droplet propagation, which suppresses the growth of the instability. 
At low enough noise, this stabilizing effect is too weak, 
the droplet evaporates, and the ordered phase is thus robust to these excitations. 

The above qualitative description stems from an analytical theory that allows us to derive an estimate for the transition line.
The resulting phase diagram is distinct from the one previously reported in the literature~\cite{Solon2015FromFluctuations}. 
Our results extend to active $O(d)$ models, 
allowing us to finally offer a comprehensive discussion
of active and passive cases.
All our numerics are detailed in~\cite{supp}.

\emph{Stability of counter-propagating droplets.}
The Vicsek model is notoriously difficult to study analytically, and we thus consider a
continuous-time, discrete-space variant in
$d=2$~\cite{solon2022susceptibility}, which recapitulates the phenomenology reported numerically in~\cite{Codina2022SmallFlock}.
On a square lattice of size $L_x\times L_y$, $N$ particles experience both diffusion biased by their continuous headings and alignment interactions. We denote the heading of
particle $i$ by ${\bf s}_i = (\cos \theta_i, \sin \theta_i)$ and its position by
$\mathbf{r}_i$. The particle hops to a neighboring site $\mathbf{r}_i+\mathbf{d} a$, where $a$ is the lattice spacing,
with rate $\frac{D}{a^2}+\frac{v}{2a}{\bf d}\cdot {\bf s}_i$, so that $D$ is a diffusivity and $v$ the
self-propulsion speed. We set $v a<2D$ to ensure positive rates. We define
the density and magnetization  at
site $\vec r_i$ as $\rho(\vec r_i)=\sum_{j\in \vec r_i} a^{-2}$ and 
$\vec m(\vec r_i)=\sum_{j\in \vec r_i}{\bf s}_j/a^2$, respectively, where the sums
run over particles at site ${\vec r}_i$. Finally, we define the polarization at site $\vec r_i$ as 
${\bf p}(\mathbf{r}_i)=\vec m(\vec r_i)/\rho(\vec r_i)$. The spin
aligns with the polarization according to the Langevin dynamics
\begin{align}
    \frac{d}{dt} \theta_i(t) = -\frac{\gamma}{T}{\bf p}(\mathbf{r}_i) \wedge \mathbf{s}_i(t) + \sqrt{2\gamma} \eta_i(t)\;,\label{eq:micro}
\end{align}
where $\eta_i(t)$ is a unit-variance Gaussian white
noise. Equation~(\ref{eq:micro}) is such that the on-site alignment dynamics corresponds to a fully-connected XY model with temperature-like parameter $T$ and $\gamma^{-1}$ the time scale of alignment. At high density and low temperatures, the system exhibits a linearly stable, collectively moving ordered state~\cite{solon2022susceptibility}. We explore the phenomenology of the system by varying the dimensionless control parameters $({\rm Pe},T,v\rho_0a /\gamma)$, where ${\rm Pe}=v^2/\gamma D$ is a 
P{\'e}clet number and $\rho_0$ the mean particle density.

To test the stability of the ordered phase against nonlinear excitations, we prepare a uniform system
at density $\rho_0$, ordered with a polarization along the $-\hat{\mathbf{x}}$ direction, and introduce a droplet of density $\rho_{\rm d}>\rho_0$ that contains particles with opposite polarization.
Above a transition temperature
$T_c({\rm Pe})$, large enough droplets propagate and grow ballistically,
destroying the ordered phase (Fig.~\ref{fig:droplets} left and
 Movie 1). Below $T_c({\rm Pe})$, on the contrary, droplets evaporate (Fig.~\ref{fig:droplets} right and Movie 2). The ordered phase of the model thus seems stable against nonlinear excitations at low enough $T$. 

 \begin{figure}
\includegraphics[width=1\columnwidth]
{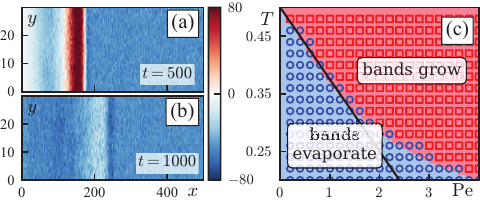}
\caption{Stability of counter-propagating bands in simulations of the active XY model. {\bf (a-b)} At $T = 0.26$ and $\mathrm{Pe} = 1.4$, a counter-propagating band inserted in the ordered phase is unstable and evaporates. 
{\bf (c)} Band stability in microscopic simulations (symbols) compared with the linear stability analysis of the leading front (colors).
The solid line corresponds to the analytical prediction for the transition line, given in
  Eq.~(\ref{eq:Tc}).
}
    \label{fig:front-stability}
\end{figure}

To make numerical and analytical progress, we work in the high density limit, $\rho_0\gg\gamma/v$, where the system is expected to order when
$T\lesssim 1/2$~\cite{solon2022susceptibility},
and we investigate the simpler
problem of the stability of the ordered phase against a dense flat band of counter-going particles spanning
the entire system in the $\hat{\mathbf{y}}$ direction, making the problem invariant along $\hat{\mathbf{y}}$. As shown in Fig.~\ref{fig:front-stability}, the band undergoes an evaporation transition akin to that of the droplets at a temperature
$T_c^b({\rm Pe})$. 
For $T>T_c^b$, the band propagates and widens ballistically as shown in Movie 3. On the contrary, for $T<T_c^b$, the
leading edge develops an instability, shown in Movie 4, that leads to the band evaporation. To account for this transition analytically, we first
show below the existence of propagating front solutions before
characterizing the transverse instability they develop below
$T_c^{b}$, and finally discuss the relationship with the loss of droplet stability at $T_c\geq T_c^b$.

\emph{Existence of propagating front solutions.}  We consider the following hydrodynamic equations for the density and magnetization fields~\cite{supp}:
\begin{align}
\!\partial_{t}\rho & =-v\bm{\nabla}\cdot\mathbf{m}+D\nabla^2\rho\;,\label{eq:rho}\\
\!\partial_{t}\mathbf{m} & =-\frac{v}{2}\bm{\nabla}\rho+D\nabla^2\mathbf{m}
+\gamma F(|\mathbf{p}|)\mathbf{m}\label{eq:m}\;,
\end{align}
where $\mathbf{p}=\mathbf{m}/\rho$ is the polarization field 
and the lattice spacing has been set to $a=1$. We consider the alignment term $F(p)=\alpha[1-(p/p_0)^2]$, where $\alpha=1/(2T)-1$ and $p_0^2=2\alpha/(1+\alpha)^2$, which amounts to a mean-field treatment of the aligning interactions~\cite{solon2022susceptibility}. 
We show in~\cite{supp} that Eqs.~\eqref{eq:rho}-\eqref{eq:m} stem from the simplest closure, 
which neglects the non-locality of the nematic field and its spatial gradients, and that our results are robust to their inclusion. 
For $T<1/2$ ($\alpha>0$), Eqs.~(\ref{eq:rho}-\ref{eq:m})
admit linearly stable uniform solutions with $|\vec p|=p_0$. 

As for the active Ising model~\cite{Benvegnen2023}, we find front solutions 
connecting regions of opposite polarity and moving at velocity $c\hat{\bf x}$, in the form
$\mathbf{m}=m_{\rm f}(z)\hat{\mathbf{x}}$ and $\rho=\rho_{\rm f}(z)$, where $z=x-ct$. Numerics show these front solutions to exist
for all ${\rm Pe}$ and $\alpha>0$, which can be proved analytically in the limit ${\rm Pe}\ll 1$~\cite{supp}. To leading order in  $\alpha\ll1$ and ${\rm Pe}\ll1$, the front profile takes the form~\cite{supp}
\begin{align}
    m_{\rm f}(z)=\rho_0 p_0 \tanh(z/w)\;\quad \rho_{\rm f}=\rho_0\;,\label{eq:dw}
\end{align}
where the speed and width of the domain wall are given by  $c\simeq v/\sqrt{2}$ and $w=(2D/\gamma\alpha)^{1/2}$. Figure~S5 in~\cite{supp} shows how the profiles obtained by numerically integrating
Eqs.~(\ref{eq:rho}-\ref{eq:m}) converge to 
Eq.~(\ref{eq:dw}) as ${\rm Pe}\to 0$.

\begin{figure}
  \centering
  \includegraphics[width=0.5\columnwidth]{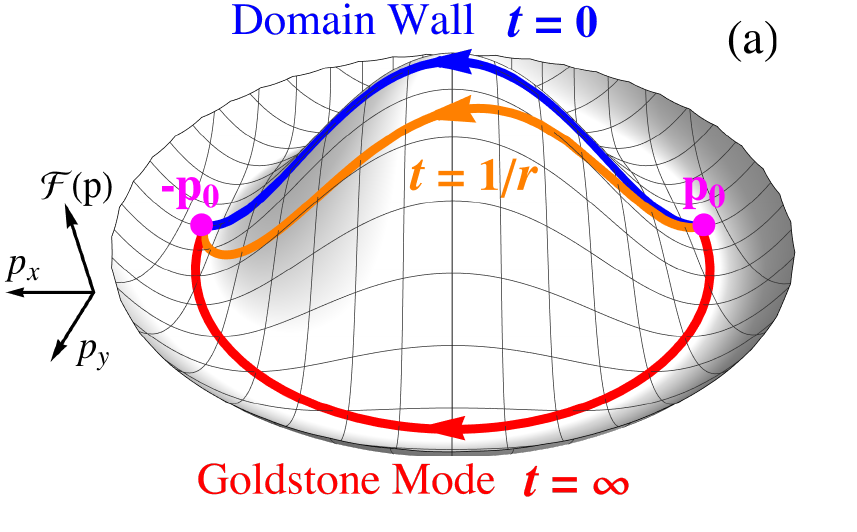}
    \includegraphics[width=0.47\columnwidth]{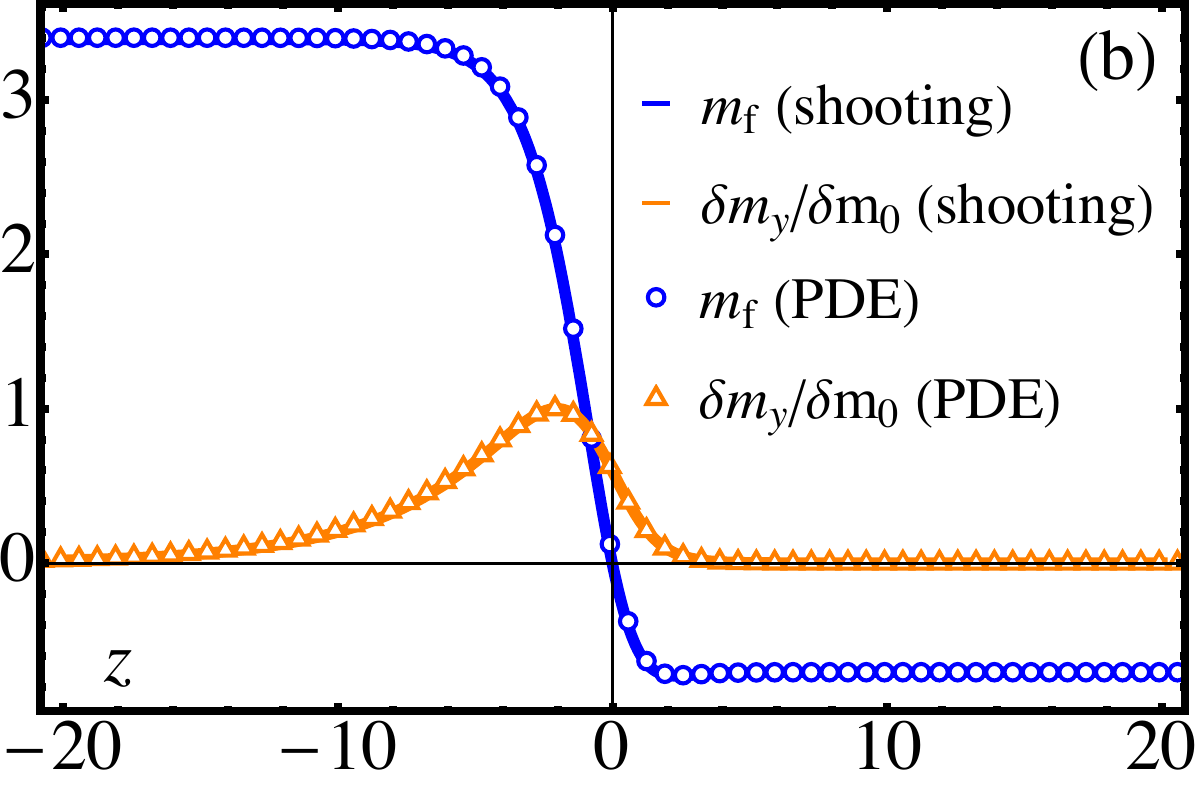}
  \includegraphics[width=0.47\linewidth]{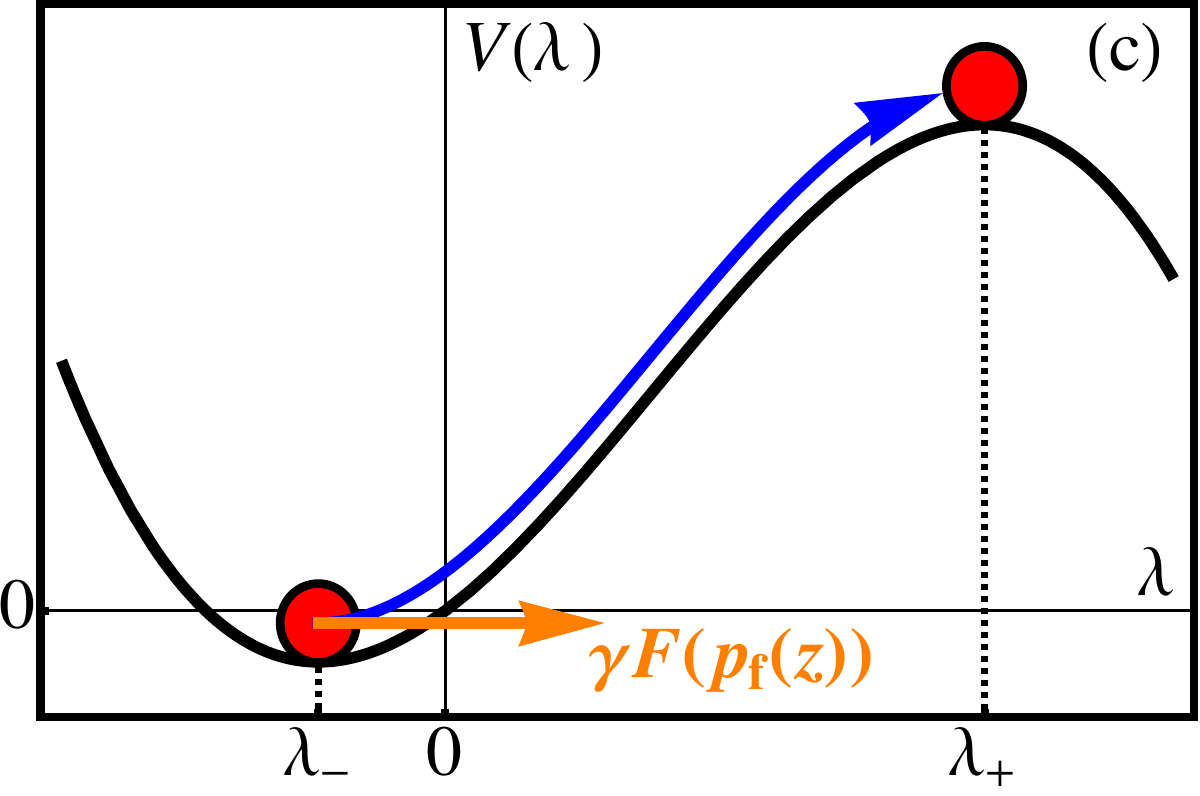}\hspace{1.5mm}
    \includegraphics[width=0.5\linewidth]{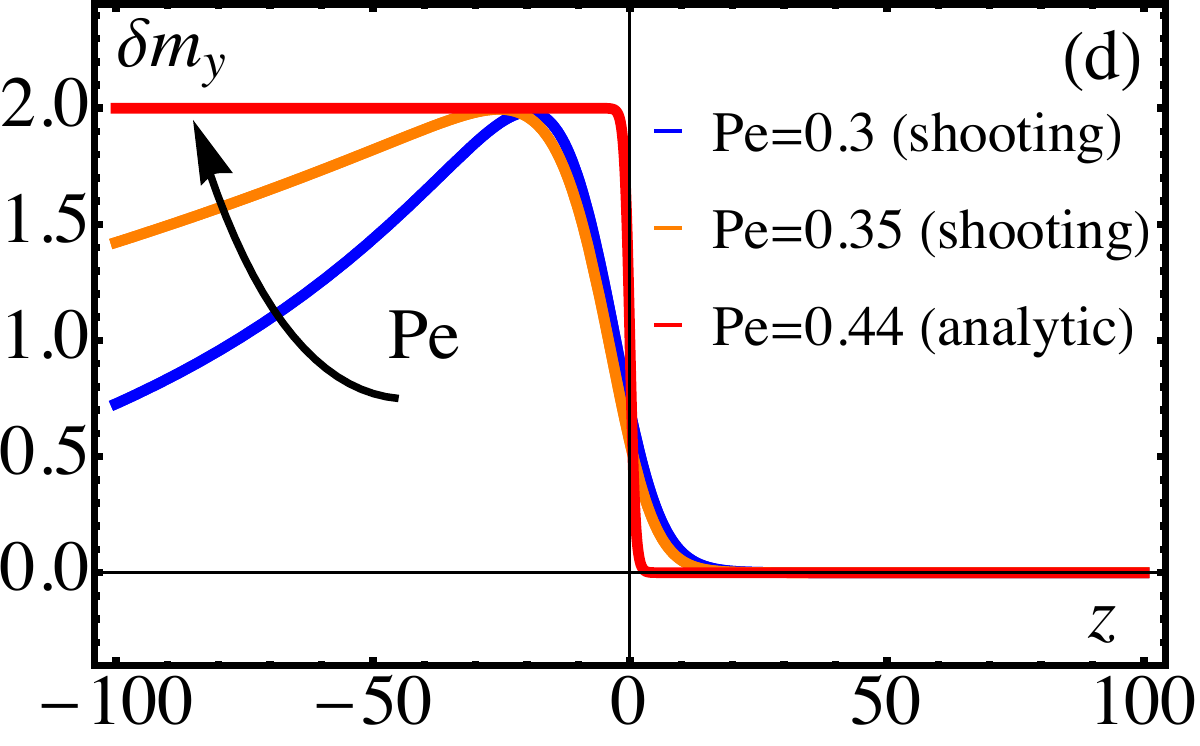}
  \caption{{\bf (a)} A domain wall (blue line) can be seen as a line under tension going through an unstable region. 
    For a passive system with $v=0$, a
    transverse fluctuation $\delta m_y/\rho$ (orange line) is
    amplified and leads to the relaxation of the domain wall into a
    Goldstone mode spanning the stable manifold $|\mathbf{p}|=|p_0|$
    (red line).  {\bf (b)} Instability profile obtained from solving
    Eq.~(\ref{eq:lambda}) by a shooting method and from direct PDE
    integration. {\bf (c)} Potential $V(\lambda)$ in
    Eq.~(\ref{eq:lambda}). There is a unique value of $r$ (the tilt
    of the potential) such that the drive $\gamma F(p_{\rm f})$ exactly
    balances the potential barrier. Parameters:
    $v=D=\gamma=\alpha=1$, giving $r=0.32$. \textbf{(d)} Instability profiles near criticality obtained via a shooting method. The critical profile (${\rm Pe}=0.44$) is partially screened and given by Eq.~\eqref{eq:inst}. (Amplitude normalized to $2$, $v=\gamma=1$, $\alpha=0.11$.)
  }
  \label{fig:newton}
\end{figure}

\emph{Stability of the front solutions.} Domain walls are intrinsically discrete-symmetry excitations: in a scalar system, opposing boundary conditions $\pm p_0$ force the polarization through $p=0$, a configuration disfavored by alignment yet topologically unavoidable. 
However, in the vectorial case, the system can develop a non-zero polarization in the transverse direction at the domain wall to remedy this frustration. 
To see this, consider the simpler case of Eq.~(\ref{eq:m}) with $v=0$ and
$\rho=\rho_0=\text{Cst}$. 
The alignment term can be written as the
derivative of a Mexican hat potential $U$, such that
$\gamma
F(|\mathbf{p}|)\mathbf{m}=-\gamma\rho\partial_{\mathbf{p}}U(|\mathbf{p}|)$
with $U(p)=-\alpha  p^2/2+\alpha  p^4/4p_0^2$. A domain wall $\mathbf{m}=m_{\rm f}(x)\hat{\mathbf{x}}$ 
connects the fixed points $\pm p_{0}\hat{\mathbf{x}}$ by going over
the top of the potential, as depicted by the blue line in
Fig.~\ref{fig:newton}a. 
The domain wall is thus linearly unstable and a small transverse
perturbation makes it relax into a Goldstone
mode that stays at the bottom of the 
potential $U$, corresponding to the red line in Fig.~\ref{fig:newton}a. 
The existence of the Goldstone mode thus generically induces the growth of transverse polarization at the domain wall.

In the presence of self-propulsion, the domain wall moves with velocity $c \hat {\bf x}$. The transverse perturbation thus has a finite time to grow before it is absorbed by the stable ordered phase once the domain wall has moved away. As we now show, the fate of the perturbation is controlled by the competition between its growth time and two time scales related to transport. 

We analyze the stability of the 
domain wall solution $[\rho_{\rm f}(x-ct),m_{\rm f}(x-ct)\hat{{\bf x}}]$ against small
transverse fluctuations $\delta m_{y}(x,t)\hat{{\bf y}}$.
Linearizing Eqs.~(\ref{eq:rho}-\ref{eq:m}) around $\rho_{\rm f}$ and $m_{\rm f}$ leads to the eigenvalue problem
\begin{equation}
\partial_{t}\delta m_{y}= D\partial_{x}^{2}\delta m_{y}+\gamma F(p_{\rm f})\delta m_{y}.\label{eq:lin}
\end{equation}
with $p_{\rm f}(z)=m_{\rm f}(z)/\rho_{\rm f}(z)$. 
We solve this equation
using $\delta m_{y}(x,t)=e^{rt}M_{y}(z)$, which leads to:
\begin{equation}\label{eq:Myofz}
rM_{y}=\left[D\partial_{z}^{2}+c\partial_{z}+\gamma F(p_{\rm f})\right]M_{y}\;.
\end{equation}
The growth rate $r$ controls the fate of the domain wall: 
if $r>0$ the domain wall is unstable, while it is stable
if $r<0$. We note that Eq.~\eqref{eq:Myofz} also describes the depinning of vortices in type-II superconductors~\cite{hatano1998}.
To determine $r$, it is convenient to parameterize $M_y(z)$ by its growth rate, $-\lambda(z)=M_y'(z)/M(z)$, using the Cole-Hopf
transformation
$M_{y}(z)=\delta m_{0}\exp[-\int_{0}^{z}dz'\lambda(z')]$, with 
$\delta m_{0}$ a constant. This simplifies Eq.~\eqref{eq:Myofz} into
\begin{equation}
D\partial_{z}\lambda=-\frac{dV}{d\lambda}+\gamma F(p_{\rm f}(z)).\label{eq:lambda}
\end{equation}
where
$V(\lambda)\equiv r\lambda +\frac{1}{2}c\lambda^{2}-\frac{1}{3}D\lambda^{3}$.

Interpreting $z$ as a time variable, Fig.~\ref{fig:newton}c shows that Eq.~(\ref{eq:lambda}) describes the overdamped dynamics of a particle at position $\lambda(z)$ in the potential $V(\lambda)$,
subject to a ``time-dependent'' nonconservative force $\gamma F(p_{\rm f}(z))$. The latter is a
positive 
pulse localized at the domain wall. 
Equation~\eqref{eq:lambda} is a joint equation for $M_y$ and $r$, which selects the
unique value of $r$ such that the potential
barrier $\Delta V\equiv V(\lambda_+)-V(\lambda_-)$, which increases monotonically with $r$, exactly matches the drive due to $F$. The solution of Eq.~(\ref{eq:lambda}) is
then a heteroclinic orbit connecting the fixed points of the
potential $\lambda_{\pm}=[c\pm\sqrt{c^2+4Dr}]/2D$. 
The resulting solution $M_y(z)$ agrees quantitatively with the growing transverse perturbation measured numerically on an unstable front, as shown in
Fig.~\ref{fig:newton}b. Here, the boundary conditions $\lambda(\pm\infty)=\lambda_\pm$ correspond to the far-field decay,
\begin{align}
M_{y} & \sim\begin{cases}
e^{-\lambda_{-}z}, & z\ll-w\\
e^{-\lambda_{+}z}, & z\gg w
\end{cases}\;,\label{eq:ff}
\end{align}
in agreement with Eq.~\eqref{eq:Myofz} when $F(p_{\rm f})=0$.

The front is unstable if $r>0$. This is equivalent to the condition that,  for $r=0$, the
particle overshoots the barrier $\Delta V$. This can be formalized
by integrating Eq.~(\ref{eq:lambda}) with $r=0$ between $z=-\infty$,
where $\lambda(z)=\lambda_-=0$, and $z=z_+$ where
$\lambda(z)=\lambda_+=c/D$. The front is then unstable if
\begin{multline}
\int_{-\infty}^{z_+}\gamma F(p_{\rm f}(z))dz=c+\int_{-\infty}^{z_+} V'(\lambda(z))dz\\<\int_{-\infty}^{+\infty}\gamma F(p_{\rm f}(z))dz\;,\label{eq:int-lambda}
\end{multline}
where we have used $F(p_{\rm f}(z))>0$ for $|z|<\infty$. 
Equation~(\ref{eq:int-lambda}) can be rewritten in terms of the competition between characteristic
time scales
\begin{align}
\tau_{{\rm inst}}^{-1}>\tau_{{\rm adv}}^{-1}+\tau_{{\rm diff}}^{-1}\;,\label{eq:cond0}
\end{align}
where $\tau_{\rm inst}=(\gamma\alpha I_{\rm inst})^{-1}$,
$\tau_{\rm adv}=w/c$, $\tau_{\rm diff}=D c^{-2} I_{\rm diff}^{-1}$ can be expressed in terms of the dimensionless functions,
\begin{equation}
u=z/w\,,\quad \Lambda(u)=\lambda(z)D/c\,,\quad\,P(u)=p_{\rm f}(z)/p_{0}\label{eq:dimvar}
\end{equation}
and integrals
\begin{align}
I_{{\rm inst}} & =\int du\left[1-P^{2}(u)\right]\;,\label{eq:tauinst}\\
I_{{\rm diff}} & =\int du\left[\Lambda(u)-\Lambda^{2}(u)\right]\Theta[1-\Lambda(u)]\;,\label{eq:taudiff}
\end{align}
where we have used $dV/d\lambda|_{r=0}=c\lambda-D\lambda^2$.

Equation~\eqref{eq:cond0} is a central result of this Letter. Up to unknown
constants of $\mathcal{O}(1)$ encapsulated in the dimensionless integrals,
it provides a general criterion for the domain wall instability, and
therefore for the stability of the ordered phase against band
excitations. Furthermore, it admits a clear physical
interpretation. The characteristic
growth time of a transverse perturbation $\delta m_y$ near the domain wall is 
$\tau_{{\rm inst}}\sim (\gamma\alpha)^{-1}$, so that $\tau_{{\rm inst}}^{-1}$ is its growth rate. However, the instability only has a finite time to grow before the front travels away due to its speed $c$. First, it takes an average time $\tau_{\rm adv}=w/c$ for the front to travel its own width, leading to an advection rate $\tau_{\rm adv}^{-1}$ of the front away from the perturbation. Additionally, particles diffuse and thus feel the advection of the front only after a time $\tau_{\rm diff}\equiv D/c^2$. The rate at which the front propagation overcomes the diffusion of the perturbation is thus $\tau_{\rm diff}^{-1}$. Equation~\eqref{eq:cond0} states that the band is unstable when the instability growth rate is larger than the total escape rate of the front, $\tau_{{\rm adv}}^{-1}+\tau_{{\rm diff}}^{-1}$.

\emph{Analytical transition line.} Let us derive an analytical prediction for the transition line, consistent with Eq.~\eqref{eq:cond0}, in the limit
$\alpha,\mathrm{Pe}\ll1$ with
$\chi\equiv\mathrm{Pe}/\alpha=\mathcal{O}(1)$. This amounts to finding the condition on $(\mathrm{Pe},T)$ such that Eq.~\eqref{eq:lambda} admits a solution for $r=0$. 
In the aforementioned limit, the domain
wall takes the form given in Eq.~(\ref{eq:dw}) so that, for $r=0$, Eqs.~(\ref{eq:lambda}) and~\eqref{eq:dimvar} lead to
\begin{equation}
\frac{d\Lambda}{du}\simeq\chi^{1/2}(-\Lambda+\Lambda^{2})+2\chi^{-1/2}[1-\tanh^2(u)]\;.\label{eq:Lambda}
\end{equation}
For the boundary conditions
$\Lambda(-\infty)=0$ and $\Lambda(+\infty)=1$, Eq.~(\ref{eq:Lambda}) admits a solution only for $\chi=4$, given by $\Lambda(u)=(1+\tanh u)/2$. Using $\alpha=1/(2T)-1$ and the definition of $\chi$, fronts are thus unstable for $T<T_c^b$, with
\begin{align}
    T^b_{\rm c}=\frac{1}{2}-\frac{1}{8}{\rm Pe}+\mathcal{O}({\rm Pe}^2)\;.\label{eq:Tc}
\end{align}
This prediction for the transition line is shown in Fig.~\ref{fig:front-stability}(a) to match
at small ${\rm Pe}$ the measurements done both in numerical
integration of Eqs.~(\ref{eq:rho})-(\ref{eq:m}) and our microscopic simulations. 
Let us now show that this prediction matches that of Eq.~\eqref{eq:cond0}. To do so, we first note that, using Eqs.~\eqref{eq:dimvar}-\eqref{eq:taudiff}, the characteristic times are given by
$\tau_{{\rm inst}}=1/(2\gamma\alpha)$ and
$\tau_{{\rm diff}}=2D/c^{2}$. Together with $\tau_{\rm adv}=w/c$, $w=2D/c$ and $c\simeq v/\sqrt{2}$, Eq.~\eqref{eq:cond0} matches Eq.~\eqref{eq:Tc} in the limit $\alpha,{\rm Pe}\ll1$, as predicted.

Furthermore, in this limit, we obtain the critical instability
profile, observed on the transition line, as
\begin{align}
\delta m_{y}(z) =\delta m_{0}\left[1-\tanh\left(\frac{z}{w}\right)\right]\;.\label{eq:inst}
\end{align}
As can be seen from Eq.~\eqref{eq:Myofz}, $\delta m_y(z)$ is a gapless excitation when $r=0$ since $F$ is localized at the domain wall. Unlike equilibrium gapless modes, which are scale-free, this one is partially screened: it decays as $e^{-c z/D}$ at large $z>0$, a feature reminiscent of forced driven-diffusive systems~\cite{Sadhu2011} and vortex depinning in type-II superconductors~\cite{hatano1998} (see Fig.~\ref{fig:newton}(d)). 
We conclude that the domain-wall instability develops when it is sufficiently slow for a gapless mode to be excited. 
This gapless mode at $r=0$ exists due to the continuous symmetry of the order parameter; it is a consequence of the Goldstone mode described in Fig.~\ref{fig:newton}(a). 
At smaller values of $T$, the spectrum is gapped and the instability $\delta m_y$ grows exponentially fast. 
All in all, the ordered phase is thus \textit{stabilized} by the existence of the Goldstone mode described in Fig.~\ref{fig:newton}(a), at odds with $2d$ equilibrium physics where it destroys order.

The reasoning above extends to dimensions $d \geq 2$. In~\cite{supp}, we show that Eq.~\eqref{eq:Tc} generalizes into
\begin{align}
    T^b_{\rm c}(d)=T_{\rm MF}(d)-\frac{\rm Pe}{2d^2(d-1)}+\mathcal{O}({\rm Pe}^2)\;,\label{eq:Tcd}
\end{align}
where $T_{\rm MF}(d)=1/d$ is the temperature below which the disordered phase is unstable. Fronts are thus unstable in all dimensions at low enough temperature. Interestingly, as $d$ increases, the region between $T_{\rm MF}(d)$ and $T_{\rm c}^b(d)$ shrinks. We attribute this to the fact that domain walls have an increasing number of transverse directions along which magnetization can grow, making them less and less stable as $d$ increases.

\begin{figure}
  \centering
  \includegraphics[width=\columnwidth]{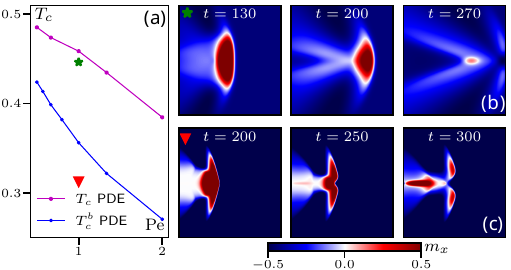}
  \caption{{\bf (a):} Comparison of the transition lines $T_c^b$ to excite a propagating band and $T_c$ to excite a propagating droplet in the hydrodynamic Eqs.~\eqref{eq:rho}-\eqref{eq:m}. The green star and red triangle
    indicate the location of the snapshots shown in the right
    panels, which illustrate the droplet instabilities: for $T<T_c^b$ (c), the center line is unstable, like band profiles, and breaks the droplet (see Movie 5), while for  $T_c^b<T<T_c$ (b) the droplet evaporates from the sides (see Movie 6). For $T>T_c$, an initial droplet grows (see Movie 7).}
  \label{fig:2d-transition}
\end{figure}

\emph{Stability of droplets.} The instability of a domain
wall necessarily implies the instability of a finite droplet for the
same parameter values, so that the criterion Eq.~(\ref{eq:cond0})
gives us a lower bound $T_c^b\le T_c$ on the temperature $T_c$ below which droplets are unstable. 
For 
the numerics shown in Fig.~\ref{fig:2d-transition}, the droplet size needed to excite a propagating solution diverges at a temperature $T_c>T_c^b$.

The analysis of the domain-wall stability described above should still apply for the tip  of macroscopic droplets, where the transverse magnetization vanishes by symmetry. The fact that $T_c>T_c^b$ suggests that a
droplet can be unstable due to an instability mechanism distinct from the one leading to Eq.~\eqref{eq:Tc}. Indeed, at
temperatures $T_c^b<T<T_c$, droplets evaporate from the sides, as
illustrated in Fig.~\ref{fig:2d-transition}b and Movie 5 while for $T<T_c^b$ the instability arises first at the center of the droplet as in Fig.~\ref{fig:2d-transition}c and Movie 6. Carrying out the stability analysis of the two-dimensional droplet solution is beyond the scope of our theory, but numerics suggests that the two
transition lines follow the same trend as a function of ${\rm Pe}$, as shown in Fig.~\ref{fig:2d-transition}a.

\emph{Discussion.} In this Letter, we have uncovered why flocking models that break a continuous symmetry have a critical dimension lower than those breaking a discrete symmetry. 
We showed that the existence of a critical mode at the tip of a counter-propagating droplet prevents it from growing at small enough noise: the presence of a gapless mode thus destabilizes
not the phase, but its nonlinear excitations.
We note that other droplet instability mechanisms may exist and, already, we identified a temperature range where droplets evaporate from the side, rather than from the center.
In the temperature range where droplets exist, the exact nature of the resulting phase is an exciting challenge, on which progress has been done recently in the discrete case~\cite{woo2024motility,Karmakar2024}.
Moreover, our work raises the need to revisit the (density, noise, motility) phase diagram of compressible flocks.
In addition, while the mechanism we uncover is generic, resulting from the competition between alignment and domain walls, we have considered only the simplest of flocking models.
In particular, constant-density ``flocks''~\cite{chen2015critical,chen2018incompressible} have been shown to be metastable due to the nucleation of very different types of defects~\cite{besse2022metastability}, whose stability would be worth studying in detail. Closer to experiments, the impact of finite compressibility due to interparticle repulsion~\cite{deseigne2010collective1,Kumar2014,Geyer2019FreezingLiquids,Soni2020} on the instability mechanisms studied here remains an open challenge.

\begin{acknowledgments}
We thank David Mukamel, Brieuc Benvegnen, and Ran Yaacoby for fruitful discussions. YK acknowledges financial support from ISF (2038/21), (3457/25) and NSF/BSF (2022605). JT thanks Laboratoire MSC for hospitality. OG acknowledges support from the Leinweber Institute for Theoretical Physics at The University of Chicago and a
MRSEC-funded Kadanoff–Rice fellowship and The University of Chicago Materials Research Science and Engineering Center, funded by NSF (DMR-2011854).
\end{acknowledgments}

\bibliographystyle{apsrev4-2}
\bibliography{./refs_new}

\end{document}